# Perfect selective alignment of nitrogen-vacancy center in diamond

Takahiro Fukui[1], Yuki Doi[1], Takehide Miyazaki[2], Yoshiyuki Miyamoto[2], Hiromitsu Kato[3,4], Tsubasa Matsumoto[3,4], Toshiharu Makino[3,4], Satoshi Yamasaki[3,4], Ryusuke Morimoto[5], Norio Tokuda[5], Mutsuko Hatano[4,6], Yuki Sakagawa[1], Hiroki Morishita[1,4], Toshiyuki Tashima[1,4], Shinji Miwa[1,4], Yoshishige Suzuki[1,4], Norikazu Mizuochi[1,4]

[1] *Graduate School of Engineering Science, Osaka University, Toyonaka, Osaka 560-8531, Japan*

[2] *Nanosystem Research Institute-National Institute of Advanced Industrial Science and Technology, Tsukuba, Ibaraki 305-8568, Japan,*

[3] *Energy Technology Research Institute-National Institute of Advanced Industrial Science and Technology, Tsukuba, Ibaraki 305-8568, Japan,*

[4] *CREST, Japan Science Technology, Kawaguchi, Saitama 332-0012, Japan*

[5] *Graduate School of Natural Science and Technology, Kanazawa University, Kanazawa 920-1192, Japan*

[6] *Department of Physical Electronics, Tokyo Institute of Technology, 2-12-1 Ookayama, Meguro-ku, Tokyo 152-8552, Japan*

**Abstract**

Nitrogen-vacancy (NV) centers in diamond have attracted significant interest because of their excellent spin and optical characteristics for quantum information and metrology. To take advantage of the characteristics, the precise control of the orientation of the N–V axis in the lattice is essential. Here we show that the orientation of more than 99 % of the NV centers can be aligned along the [111]-axis by CVD homoepitaxial growth on (111)-substrates. We also discuss about mechanisms of the alignment. Our result enables a fourfold improvement in magnetic-field sensitivity and opens new avenues to the optimum design of NV center devices.

One of the most intensively studied atom-like solid-state systems is the nitrogen-vacancy (NV) center in diamond. It is composed of a substitutional nitrogen (N) and a vacancy (V) on adjacent lattice sites in diamond (Fig. 1). It has attracted significant interest because of its excellent spin and optical characteristics for quantum information processing, communication, sensing and metrology.[1-6]

In the diamond crystal structure, the orientations of NV centers are classified according to the alignment along one of four possible crystallographic axes: [111], [1$\bar{1}\bar{1}$], [$\bar{1}$1$\bar{1}$], or [$\bar{1}\bar{1}$1] (Fig. 1). In most diamond samples, NV centers equally occupy these four orientations. It was recently shown that in the case of synthetic diamond grown via chemical vapor deposition (CVD) on (110) substrates, NV centers can be incorporated into the lattice and found in only two of four orientations.[7] Furthermore, this 50% preferential NV orientation can be realized in CVD diamond samples grown on (100) substrates.[8] In this sample with NVs oriented only along two axes, twofold improvement in readout contrast ($R$) and a magnetic field sensitivity ($\eta$) were demonstrated when compared to with those of standard samples with equal population of all NV orientations.[8] It is consistent with the relationship in shot-noise limited sensitivity of $\eta \propto 1/R$.[6,8] It also indicates that fourfold improvement in $\eta$ can be expected if the orientation is aligned in only one orientation.

However, such a perfect alignment has not yet been realized. Control of the orientation of the N-V axis is very important for not only sensing but also quantum information applications because the spin and optical characteristics strongly depend on this orientation. In cases where photoluminescence (PL) is detected from the [111] direction, the PL intensity from N–V centers in which the N–V axis is parallel to [111] (NV ∥ [111]) is higher than others because electric dipole transitions are allowed for dipoles in the plane perpendicular to the N–V axis.[9] Furthermore, since growth is performed on a (111) face, the magnetic field can be directly aligned perpendicularly to the surface, which is the easiest configuration from a practical point of view. With respect to spin, the electron spin of negatively charged NV (NV⁻) is expected to play a key role at the quantum interface with photon[1] and superconducting flux qubits.[2] The coupling with nearby NV⁻ is also expected to be a resource of quantum register.[3] In future applications of these systems, precise control of not only the position but also the orientation of the N–V axis will be required.

We investigated five samples (Table 1). The conditions used for diamond growth or NV centers creation differed among them. The diamond films of samples A, B and D were epitaxially deposited on high-pressure and high-temperature synthetic (HPHT) Ib diamond (111) substrates (2 mm × 2 mm × 0. 5 mm in A, B and 1 mm × 1 mm × 0. 3 mm in D). These samples were grown in a microwave plasma-enhanced (MP) CVD reactor (ASTeX) using 1% $CH_4$ diluted with $H_2$ in cases A, B and in a reactor (Arios, spherical resonant cavity type)[10] using 0.25% $CH_4$ diluted with $H_2$ in D. The gas pressure, total gas flow rate and microwave power were 150 Torr, 1,000 sccm and 3,500 W, respectively, for A; 130 Torr, 990 sccm and 3,700 W, respectively, for B; 20 kPa, and 100 sccm and 400 W, respectively, for D. In B, $O_2$ was added (0.5 sccm) to suppress the incorporation of

impurities.[Kar] The substrate temperature was maintained at 850°C, 1,100°C and 900°C for A, B and D, respectively. The film thicknesses of A, B and D were approximately 11, 8, and 18 μm, respectively. N was unintentionally incorporated during the growth of A, B and D. Because no NV centers were observed in B and C, $^{15}$N was ion implanted into these samples B and C (IIa HPHT (111) substrate, 2 mm × 2 mm × 0. 5 mm) with a dose of $1\times10^9$/cm$^2$ and at an acceleration energy of 30 keV at 600 °C. In sample E, electrons were irradiated with 0.5 MeV with a density of $1.5\times10^{16}$/cm$^2$ to HPHT Ib(111) diamond. Samples B, C, and E were annealed at 1,000 ºC under an Ar atmosphere for 2 h and were subsequently cleaned with H$_2$SO$_4$ and HNO$_3$ solutions at 200 °C for 30 min.

The orientation of the N–V axis was investigated by optically detected magnetic resonance (ODMR) using a confocal microscope at room temperature. The electron ground state of NV$^-$ is a spin triplet. Upon optical excitation, NV$^-$ exhibits strong fluorescence. The fluorescence intensity of NV$^-$ is spin dependent owing to spin-selective relaxation via a singlet state, which allows optical read out of the single-electron spin resonances of $|M_S\rangle = |0\rangle \Leftrightarrow |-1\rangle$ and $|M_S\rangle = |0\rangle \Leftrightarrow |1\rangle$ transitions. Microwave fields are used for coherent manipulation of a single-electron spin. An external magnetic field was applied by a permanent magnet.

To quantitatively characterize the electron spin states, the ODMR spectra were simulated by exact diagonalization of the spin Hamiltonian.

$\mathbf{H} = g_e\beta_e\tilde{\mathbf{S}} \cdot \mathbf{B} + \tilde{\mathbf{S}} \cdot \mathbf{D} \cdot \mathbf{S},$

where the electron spin $S = 1$ is considered and $\beta_e$ is the Bohr magneton. Reported values for the zero-field splitting parameter ($D$=2.87 GHz) and the isotropic electron Zeeman g-value ($g_e$=2.0028) were used.[12] The ODMR spectrum of NV$^-$ can be mainly explained by Zeeman splitting ($g_e\beta_e\tilde{\mathbf{S}} \cdot \mathbf{B}$) and zero-field splitting ($\tilde{\mathbf{S}} \cdot \mathbf{D} \cdot \mathbf{S}$) if hyperfine splitting can be neglected.[13] We assume that the zero-field splitting parameter of $E$ is zero because the symmetry of NV$^-$ belongs to $C_{3v}$ point group.[13] Under this condition and a constant magnetic field strength, the ODMR frequency is simply determined by the angle between the directions of the N–V axis and the magnetic field ($\theta_{NV-B}$, Fig. 1).[13] As seen from Fig. 1, four $\theta_{NV-B}$s angles are possible depending on the four orientations of the N–V axis under an arbitrary orientation of the magnetic field, which produces four ODMR frequencies.

A scanning confocal microscope image of sample A is shown in Fig. 2(a). Bright spots are single NV$^-$, which were confirmed by the PL spectrum (Fig. 2(e)) and anti-bunching measurements. An ODMR spectrum of a single NV$^-$ in sample A is shown in Fig. 2(b). The magnetic field was applied along the [111] axis (**B** ∥ [111]) with a magnitude of 5.5 mT. The $|M_S\rangle = |0\rangle \Leftrightarrow |-1\rangle$ and $|M_S\rangle = |0\rangle \Leftrightarrow |1\rangle$ transitions were observed at low and high frequencies, respectively. The difference in their ODMR intensities is considered to result from the impedance mismatching of a Cu wire for irradiation by a microwave field. A simulated spectrum of a single NV$^-$ in the case where the magnetic field is parallel with the N–V axis ($\theta_{NV-B} = 0°$) is shown in Fig. 2(c); the simulated spectrum well reproduced the experimental results. The simulated spectrum of a single NV$^-$ in the case where $\theta_{NV-B} = 109.47°$ is shown in Fig. 2(d). This corresponds to the situation where $\mathbf{B} \parallel [111]$ and $\mathrm{NV} \parallel [1\bar{1}\bar{1}], [\bar{1}1\bar{1}], [\bar{1}\bar{1}1]$. In this case, the ODMR frequencies of $\mathrm{NV}^-$ of $\mathrm{NV} \parallel [1\bar{1}\bar{1}], [\bar{1}1\bar{1}], [\bar{1}\bar{1}1]$ corresponds to each other because of the $C_{3v}$ symmetry; however, these frequencies differ from those of $\mathrm{NV} \parallel [111]$ as shown in Figs. 2(c, d). Therefore, we can clearly identify the orientation of the N–V axis from the ODMR spectrum. In total, we investigates 100 randomly chosen single $\mathrm{NV}^-$ in sample A and found that all orientations of the N–V axes were parallel to the [111] axis. On the basis of these results, the orientations of more than 99 % of the N–V axes were aligned along the [111] axis in sample A. In samples B and C, 100 and 50 randomly chosen single $\mathrm{NV}^-$ centers, respectively, were measured. The results indicate that about 43% and 35% of the $\mathrm{NVs}^-$ exhibit $\mathrm{NV} \parallel [111]$ alignment in samples B and C, respectively, as summarized in Table 1. It should be noted that we can not distinguish whether $\mathrm{NV} \parallel [111]$ or $\mathrm{NV} \parallel [\bar{1}\bar{1}\bar{1}]$ in our experiment.

The $\mathrm{NV}^-$ concentration in sample D is higher than that in sample A; we measured $\mathrm{NV}^-$ ensembles. The difference in the NV densities between samples A, B and D is considered to be due to the difference of background pressures of $1\sim2\times10^{-8}$ Torr in the reactor for the sample A and B and $10^{-6}$ Torr for the sample D. In addition, $O_2$ was added for sample B. Notably, both samples A and B were synthesized under high-quality synthesis conditions that favored lateral growth.[14] On the basis of the approximately tenfold greater PL intensity of $\mathrm{NV}^-$ compared with that of a single $\mathrm{NV}^-$, we surmised that the ODMR spectrum of sample D in Fig. 3(a) consists of signals of approximately 10 $\mathrm{NVs}^-$. For reference, the ODMR spectrum of sample E (red) under the same direction of magnetic field with Fig. 3(a) (almost $\mathbf{B} \parallel [111]$) and its simulated spectrum (blue) are shown in Fig. 3(b). These consist of $\mathrm{NV} \parallel [111]$ and $\mathrm{NV} \parallel [1\bar{1}\bar{1}], [\bar{1}1\bar{1}], [\bar{1}\bar{1}1]$ signals. In the simulation, we slightly tilted the direction of the magnetic field to reproduce well the experimental results because the direction of the magnetic field slightly deviated from the [111] axis in our measurements. The definitions of angles $\theta_B$ and $\phi_B$ are shown in Fig. 3(d). The parameters used in the simulation of the spectrum in Fig. 3(b) were a magnetic field of 5.05 mT, $\theta_B = 3°$, and $\phi_B = 6°$. A comparison of the spectra in Figs. 3(a,b) reveals that all of the NV in sample D exhibit a $\mathrm{NV} \parallel [111]$ orientation. We investigated more than 10 randomly chosen locations and confirmed that all of the NV exhibit a $\mathrm{NV} \parallel [111]$ orientation. On the basis of this result and the simultaneous measurement of approximately 10 NVs in our detection spot, we estimate that more than 99 % of the NV in sample D exhibit a $\mathrm{NV} \parallel [111]$ orientation.

Fig. 3 (c) shows ODMR spectrum of sample E under almost $\mathbf{B} \parallel [1\bar{1}\bar{1}]$ (red) and a simulation spectrum (blue) are shown. The spectra simulated for a ratio of about 43% $\mathrm{NV} \parallel [111]$ exhibited the best fit, as shown by the blue lines in Figs. 3(b, c). The parameters used to simulate the spectrum in Fig. 3(c) were a magnetic field of 4.5mT, $\theta_B = 109.4°$, and $\phi_B = 8°$. In Fig. 3(a-c), the signal intensities were normalized with their integrated signal intensities. In samples B, C and E, the NV centers were slightly preferentially aligned with the [111] axis. We deduce that the reason relates to the fact that implantation and e-irradiation were performed from the direction of the [111] axis. As far as we know, results related to the preferential alignment by ion-implantation and e-irradiation have not yet been reported; elucidation of the mechanism and improvement of the ratios are very important in the next step.

Hereafter we perform a theoretical investigation as to how the NV centers are aligned in the <111> direction when they are created. It should be reminded that we are not discussing whether the NV centers are more favorable to form than the single substitutional N defects. The most probable scheme of NV formation seems stepwise, as Atumi et al.[15] noticed, where the nitrogen atom is incorporated first, and the vacancy second. The N should be three-fold coordinated with the lone pair as a source of vacancy. During CVD crystal growth the C ad-atoms should "skip" the N lone pairs, because they tend to stick to the C dangling bonds to gain as much energy per unit time as possible. Our first-principles energetics[16] shows that putting a $CH_3$ unit on N substituting for the topmost C atom in a flat (111) terrace is large (3 eV) relative to the reference with $CH_3$ on the topmost C itself.[17] If this type of N atoms are abundant the NV centers should be [111]-oriented with additional crystal layers grown over.

Meanwhile, one cannot exclude formation of NV defects oriented in $[1\bar{1}\bar{1}], [\bar{1}1\bar{1}],$ and $[\bar{1}\bar{1}1]$, depending on formation energetics of N atoms in different environments from the topmost terrace sites. Therefore, in order to explain the experiment, it is necessary to find in the C(111) growth process an origin of the alignment which allows us to generate the N atoms in the topmost C positions of the surface.

The lateral layer-by-layer growth is essential to a high quality CVD growth of diamond surfaces.[18,19] Edmonds et al.[7] discuss the origin of the NV alignment either in the [111] or $[\bar{1}\bar{1}1]$ direction in the case of the (110) surface. They argue that, once an N atom is incorporated into the trough flow, the site adjacent to the N atom is very attractive for C addition since the chemisorbed atom would form two bonds with the surface, while the N atom adopts a threefold coordination.

A recent experiment for a high quality layer-by-layer growth of the (111) surface has demonstrated that the lateral growth results from "kink flow", where the kink (Fig. 4(a)) rapidly propagates along the step edge taking ad-atoms in it. If the kink runs from one apex of the island to the other, then the step edge grows by a unit length in the step down direction.[20]

Owing to this growth picture, we further performed first-principles energetics[16] of various structures with N in kinks at the step edge of C(111)-1x1:H. The configuration leading to the [111]-oriented NV centers is that shown in Fig. 4 (a) in which the N occupies the α site rather than the one with N at the β site of the kink.

We indeed identified possible atomic structures that favor the N in the α position at the kink of the $[\bar{1}\bar{1}2]$ step (Fig. 4(a)). This step has been studied not only theoretically for the hydrocarbon incorporation mechanism[21,22] but also experimentally observed.[20,23] We considered two cases with respect to the N position at the kink (α and β sites), each of which is further examined in two different step-edge structures (reconstructed and unreconstructed). Then we found that the N at the α kink site is energetically favorable to form for both reconstructed and unreconstructed step edges. The energy gain relative to N at the β site is 0.51 eV for the reconstructed step edge (Fig. 4(b) relative to Fig. 4(c)) and 0.46 eV for the unreconstructed step edge (Fig. 4(d) relative to Fig. 4(e)). Combining with the consideration above, our result shows that the layer-by-layer growth via the fast kink flow enables the [111]-oriented NV centers to be frozen in the (111) surface in the absence of unwanted NVs arranged in other directions. Thus our theoretical study explains the experiment.

In summary, ODMR experiments revealed that the orientations of more than 99 % of the NV can be aligned along the [111]-axis in diamond through high-quality CVD homoepitaxial growth on (111) substrates. A possible clue to the NV alignment in lateral layer-by-layer growth is preferential generation of N atoms occupying the topmost C positions at the kinks in the $[\bar{1}\bar{1}2]$ steps, which will become three-fold coordinated after the surface-layer completion. Their lone pairs produce the vacancies on top of N and the [111]-aligned NV centers are embedded in the subsequent growth.

While preparing this manuscript, we became aware of similar researches submitted in arXiv related to the alignment of NV in diamond grown on diamond (111) substrates by the CVD growth technique.[24,25] In one of the submitted preprints, the authors[24] briefly comment on a possible mechanism of the NV alignment focusing on the step equivalent to the $[11\bar{2}]$-oriented one as shown in Fig.4 (a).

**Acknowledgement**

The authors acknowledge financial support by SCOPE, JST-CREST, KAKENHI, and NICT programs.

**Fig. 1**, Four possible orientations of NV centers in diamond. The green and brown arrows indicate the orientations of the N–V axes in a magnetic field. The angle between them is represented by $\theta_{\text{NV-B}}$.

**Fig. 2** (a) Scanning confocal microscope image of sample A. The excitation wavelength was 532 nm, and the laser power was 200 μW. The bright spots are single NV⁻ which were confirmed by PL and anti-bunching measurements. (b) The ODMR spectra of single NV in sample A. The magnetic field was applied along the [111] axis (**B** ∥ [111]). The $|M_S\rangle = |0\rangle \Leftrightarrow |-1\rangle$ and $|M_S\rangle = |0\rangle \Leftrightarrow |1\rangle$ transitions are observed at low and high frequencies, respectively. (c) The simulated spectrum of a single NV in the case of **B** ∥ NV ($\theta_{\text{NV-B}} = 0°$). (d) The simulated spectrum of a single NV in the case of $\theta_{\text{NV-B}} = 109.47°$. This corresponds to the situation where **B** ∥ [111] and NV ∥ [1$\bar{1}\bar{1}$], [$\bar{1}$1$\bar{1}$], [$\bar{1}\bar{1}$1]. (e) A PL spectrum of a single NV⁻. The excitation wavelength was 532 nm, and the laser power was 100 μW.

**Fig. 3** (a) ODMR spectrum of ensemble NV centers in sample D under almost **B** ∥ [111]. (b) ODMR spectrum of sample E at almost **B** ∥ [111] (red) and simulated spectrum (blue). This consists of the signals of NV ∥ [111] and NV ∥ [1$\bar{1}\bar{1}$], [$\bar{1}$1$\bar{1}$], [$\bar{1}\bar{1}$1]. The parameters were a magnetic field of 5.05mT, $\theta_B = 3°$, and $\phi_B = 6°$. (c) ODMR spectrum of sample E with almost **B** ∥ [1$\bar{1}\bar{1}$] (red) and simulated spectrum (blue). The parameters used in simulation were a magnetic field of 4.5mT, $\theta_B = 103°$, and $\phi_B = 4°$.

**Fig.4** (a) Schematic representation of a kinked [$\bar{1}\bar{1}$2]- and a straight [11$\bar{2}$]-oriented steps of the (111) diamond surface. We focus on the former in this paper. Green and yellow circles represent the atoms in two sublattices in one bi-layer, respectively. The former are the topmost C atoms while the latter are the second-top C atoms. The kink sites are labeled α and β. Panels (b), (c), (d) and (e) represent optimized structures of a nitrogen atom at the kink of the [$\bar{1}\bar{1}$2] step edge. In each panel, a dashed zigzag line is a guide to the eyes to help identify the position of the kink. Circles colored with grey, blue and red represent C, H, and N atoms, respectively. Panels (b) and (c): N atoms at the α and β kink sites, respectively, where both step edges are reconstructed. Panels (d) and (e): N atoms at the α and β kink sites, respectively, where both step edges are unreconstructed.

**Table 1,** The ratio that the N–V axis is aligned along with [111] in each samples.

| Sample | Synthesis | Incorporation of NV | Measured NV | NV ∥[111] | Thickness of CVD film |
|---|---|---|---|---|---|
| A | CVD growth on Ib(111) substrate by ASTeX reactor | grow-in | Single | > 99 % | 10 μm |
| B | CVD growth on Ib(111) substrate by ASTeX reactor | $^{15}$N ion-implantation (Intentional addition of O$_2$) | Single | 42% | 8 μm |
| C | IIa HPHT | $^{15}$N ion-implantation to (111) surface | Single | 35% | |
| D | CVD growth on Ib(111) substrate by Arios reactor | grow-in | Ensemble | > 99 % | 24 μm |
| E | Ib HPHT | e-irradiation to (111) surface | Ensemble | 43% | |

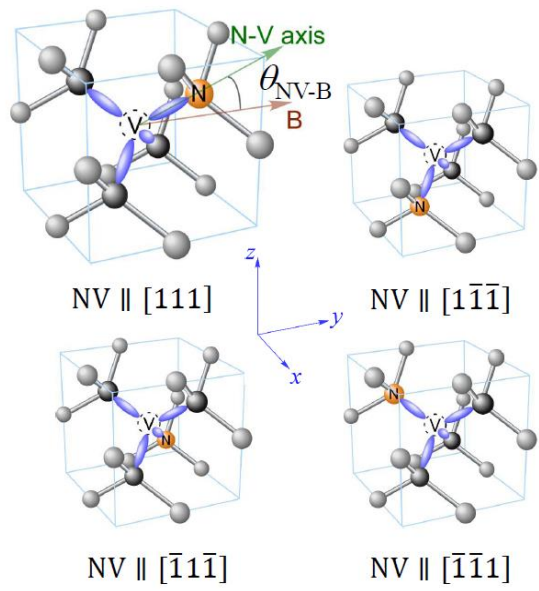

Fukui et al., Fig. 1

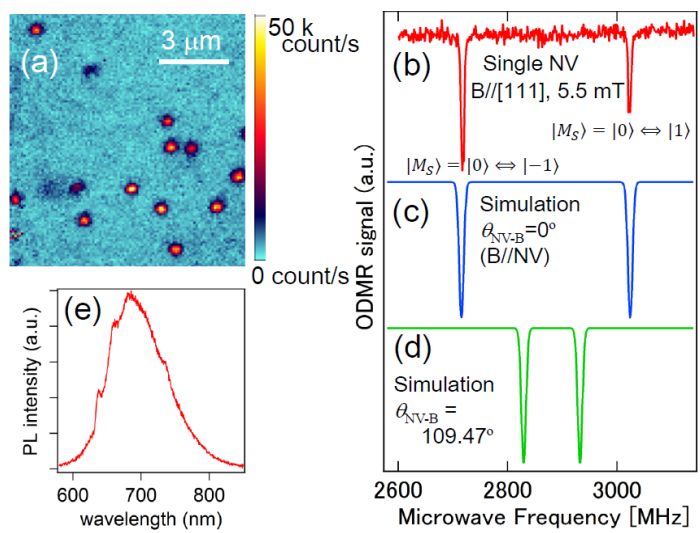

Fukui et al., Fig. 2

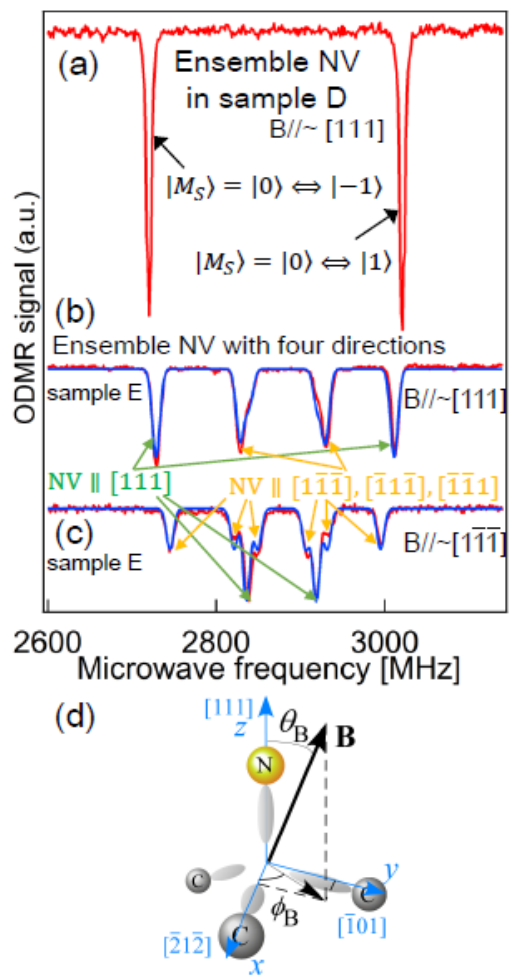

Fukui et al., Fig. 3

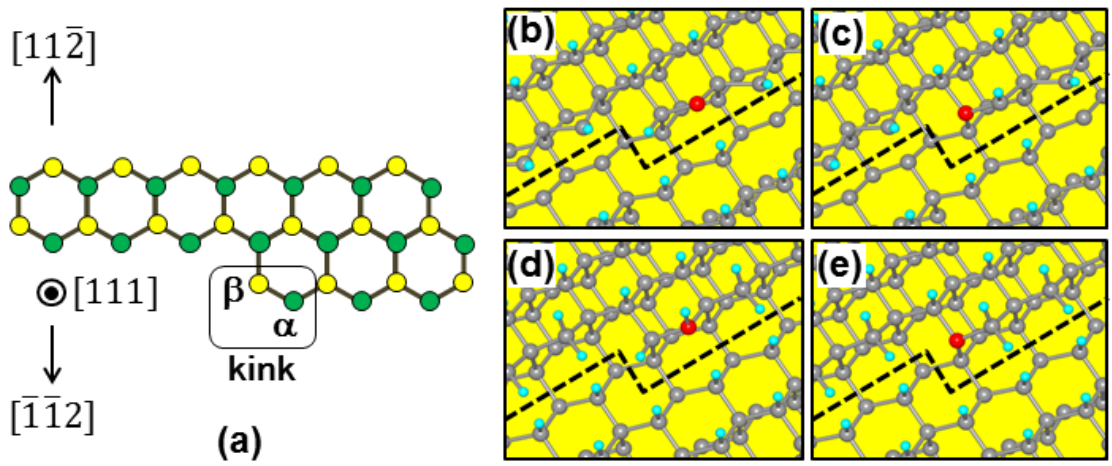

Fukui et al., Fig. 4